\documentclass{llncs} 
\usepackage{ucs} 
\usepackage[utf8x]{inputenc} 
\usepackage{url} 
\usepackage[bookmarks=false]{hyperref}
\usepackage{graphicx}
\usepackage{subfigure}
\usepackage{multirow}
\begin{document} 
\title{Increasing GP Computing Power via Volunteer Computing}
\author{Daniel Lombraña González\inst{1} \and Francisco Fernández de Vega\inst{1} \and L.
Trujillo\inst{2} \and G. Olague \inst{2}\and F. Chávez de la O\inst{1} \and M. Cárdenas\inst{3} \and L. Araujo\inst{4} \and P.
Castillo\inst{5} \and K. Sharman\inst{6}}
\institute{University of Extremadura \email daniellg@unex.es, \email fcofdez@unex.es, \email fchavez@unex.es
\and CICESE \email trujillo@cicese.mx, \email olague@cicese.mx
\and Ceta-Ciemat \email miguel.cardenas@ciemat.es
\and UNED \email lurdes@lsi.uned.es
\and University of Granada \email pedro@atc.ugr.es
\and University Politécnica of Valencia, \email ken@iti.upv.es}
\maketitle

\begin{abstract}
    This paper describes how it is possible to increase GP Computing Power via Volunteer Computing (VC) using the BOINC framework. Two experiments
    using well-known GP tools -Lil-gp \& ECJ- are performed in order to demonstrate the benefit of using VC in terms of computing power
    and speed up. Finally we present an extension of the model where any GP tool or framework can be used inside BOINC regardless of its
    programming language, complexity or required operating system.
\end{abstract}

\section{Introduction}
\label{introduction}
Real world optimization problems are usually complex and their resolution is CPU time consuming when EAs are employed. This is due to the big
amount of individuals which are evaluated and also due to the number of iterations required to find a solution.

In order to alleviate this problem, EAs and GP have benefited from parallel models. 
Two main approaches to parallelize have been described: (see \cite{spatially-structured-EAs}): \emph{Fine-grain} which uses a
master-slave architecture and \emph{Coarse-grain} also known as island model.
Given the stochastic nature of EAs and GP and the large number of runs, frequently required for obtaining results, parameter sweep
models have recently been applied in combination with high throughput computer systems. 

GRID computing is nowadays one of the most emerging technologies and has been recently
demonstrated as a powerful tool to deal with time-consuming applications, see \cite{the-grid}. The GRID allows  
super computers, clusters or desktop PCs, which are distributed over networks, to be harnessed by means of a special software called middleware. 
The middleware exports and handles all the computer resources with the goal of providing a standard layer where scientists can run their experiments.

The middleware can be focused on handling two type of resources: desktop PCs or super computers/clusters. An example of super
computer-cluster middleware is gLite (see \cite{glite}). A successful attempt of using EAs and GRID computing 
is presented by N. Mealab et al. see \cite{grid-parallel-bioinspired-algorithms}. However, this kind of middleware is usually
associated with expensive hardware which is a drawback for scientists from developing regions or countries. 

There exists other middleware systems which focus on cheap desktop PCs. This type of middleware is aimed at building up
Desktop Grid Computing (DGC) systems. The complexity and deployment of this technology is much smaller than alternative GRID one
(such as gLite). There are several DGC systems such as: 
\emph{Xtremweb} \cite{xtremweb} a research project which presents a Global Computing platform using a 
large base of volunteer PCs, \emph{Condor} \cite{condor} a middleware which implements a scheduling system with desktop PCs and
\emph{BOINC} \cite{boinc-paper} a multi-platform middleware that uses workstation CPU idle periods to run jobs. 

When dealing with DGC, the users are really important. DGC relies on them because the idea behind DGC is to allow users to collaborate with a scientific
research project by donating desktop computing power. To the best of our knowledge, the pioneer on engaging users to collaborate with a scientific research was the project 
SETI@HOME \cite{setiathome}. This project has been able to create a super virtual computer of 259.729 TeraFLOPS thanks to the collaboration
of 706,586 users\footnote{Data obtained from the web \url{http://boincstats.com}}. Thus, DGC is also known as Volunteer Grid Computing (VGC) due to 
the users that unselfishly collaborate with a scientific research. Therefore, VGC computing power can be used for free if users see the
interest of the project. 

From all the above presented VGC middleware technologies BOINC is the most used one. Furthermore, BOINC is already widely used in different
research fields such as: \emph{High Energy Physics} \cite{lhc}, \emph{Astrophysics} \cite{einsteinathome}, \emph{Climate Prediction}
\cite{climateprediction}, \emph{Epidemiology} \cite{africaathome}, etc.

A novel technology has also been recently described which allows to harness computing resources by only browsing a given and special
web page \cite{stealth-computing0,stealth-computing,stealthcomputing2}. Yet, we don't consider it here because it allows to avoid users to
notice the background work of the web-browser, thus computing power is ``stolen'' from user PCs. Users will thus be annoyed if they
discovered that someone has been using their resources without their permission. Our idea is just the opposite, not only to inform
users about the project that we are running but also invite them to join and collaborate. 

VGC is also a good computing platform for running Parameter Sweep experiments in genetic and evolutionary computation. M.E.
Samples describes Commander, a new generic parameter sweep framework (see \cite{parameter-sweep-gp}). However, firstly, is not aimed at
harnessing  volunteer resources, secondly, has not been so widely spread and tested with real scientific research problems, and thirdly
it hasn't involved the huge number of clients like BOINC (2,364,170 computer clients), a consolidated VGC.

Any of the above parallel described models -fine and coarse grain- can run within this approach. We simply have to bear in mind that any of the whole parallel
models will run on a single machine. The power comes from the multiple and simultaneously runs of the same experiment with different
parameters or identical runs for statistical analysis. 

Therefore, what we propose is to use a VGC BOINC model and GP in order to:
\begin{itemize}
    \item Harness a large number of BOINC resources (nowadays BOINC has 2,364,170 computers in total which provides in average
        668,541.2 GigaFLOPS\footnote{Data obtained from \url{http://boincstats.com under} BOINC Combined stats}).
    \item Improve the speed up of GP sequential executions thanks to the parallel environment which VGC provides. 
\end{itemize}

We have chosen BOINC as our VGC middleware because BOINC is the most used one, therefore allowing a great computing power.

The remainder of the paper includes a description of the BOINC model in Section \ref{boinc-model}; we present
the new VGC and GP model in Section \ref{vgc-gp}; Section \ref{experiments} shows the experiments and results and an extension
to the model. We conclude in Section \ref{conclusions}.

\section{The BOINC model}
\label{boinc-model}
As described above, BOINC is a middleware that harness the commodity computer resources for a given project. BOINC has two main
key features: it's \emph{Multiplatform} and \emph{Open source}. BOINC uses a master-slave model where the server is in charge of:
\begin{itemize}
    \item \emph{Hosting the scientific project experiments}. One project is composed by a binary (the algorithm) and some input files. 
        The  binary is classified according to the target platform (Ms. Windows, GNU/Linux, MacOSX) and architecture (x86 32-64 bits 
        and sparc). 
    \item \emph{Creation and distribution of jobs}. In BOINC's terminology a job is called ``work unit'' (WU). A WU describes how the
        experiment must be run by the clients (the name of the binary, the input/output files and the command line arguments).
    \item \emph{Assimilation and validation of results}. When the clients finish the computations, they upload the results to the server.
        At this point the server runs two processes: a validation and assimilator program. The validator goal is to verify if the results
        are corrupted or not. If everything is correct the results are validated and prepared for the next program: the assimilator. The
        assimilator is in charge of parsing the output files of the project to perform tasks like: compute some statistics, store results
        inside other database, etc. 
\end{itemize}

The client is quite simple. The BOINC client connects to the server and asks for work (WU). The client downloads the necessary files
and starts the computations. Once the results are obtained, the client uploads them to the server. Additionally, during the whole process 
(all the steps: download WU, process it, upload the results) the client contacts the server to keep a ``heartbeat'' function.
The heartbeat is used to take some decisions and generate some statistical data. 

As BOINC relies on users, BOINC resources are not reliable because:
\begin{itemize}
    \item \emph{Users turn on and off its machines without knowing if they are interrupting a BOINC execution}. Therefore, the research
        application must have a checkpoint facility.
    \item \emph{Users could try to cheat}. BOINC has a redundancy feature that circumvents this problem. The BOINC administrator can
        define which is the minimum required quorum to validate a solution.
    \item \emph{Users could try hacking the BOINC server}. If one user could hack the server, he could be able to launch new WUs which can
        have viruses. In order to avoid this problem, BOINC uses digital signatures to sign binary applications. Therefore, only signed applications can be
        distributed over the clients.
\end{itemize}

As we can observe BOINC's structure is simple and provides the main needed features that any VGC system requires. The next sub-section explains how we can use 
BOINC with a scientific research project.

\subsection{How to use BOINC with a Scientific Project}
\label{howto-use-boinc}
A scientific project that wants to use BOINC has to set up a GNU/Linux server (Apache, MySQL, PHP) and then take into account the
following key points:
\begin{itemize}
    \item \emph{Programming Language}. BOINC uses C++ and Fortran. Thus, if the scientific project is not coded in C++ or FORTRAN the
        project has to be ported to C++ or Fortran in order to link its source code with the BOINC framework.
    \item \emph{Operating System}. BOINC has clients for GNU/Linux, Ms. Windows and MacOSX. However, the scientific project has to be
        adapted to all of them if we want to harness as much as possible available resources. BOINC uses a generic framework which
        builds binaries which are OS dependant.
\end{itemize}

Therefore, there are two ways to support BOINC:
\begin{itemize}
    \item \textbf{Method 1. }\emph{To port the code}. This method is the most used. Basically, a researcher has to adapt its application source code to
        support BOINC. The changes could be easy if the tool is coded in C/C++ or Fortran. In other cases the research will have to
        rewrite the whole code.
    \item \textbf{Method 2. }\emph{The Wrapper}. The BOINC team provides a tool called \emph{wrapper} which enables to run
        statically linked software inside BOINC without needing to modify or port the application source code. Basically the wrapper embodies the application in such
        a way that for the BOINC client does only exists one application: the wrapper. 
\end{itemize}

In conclusion, an application which is not coded in C/C++/Fortran will use the second method. However, if the application is coded
in C/C++/Fortran some minor changes will be needed to support BOINC (\textbf{Method 1}). 

\section{VGC and GP problems}
\label{vgc-gp}
Our proposal is to use VGC for running GP experiments via BOINC technology. We present two examples that show how any GP preferred tool
could be effectively used within the BOINC framework. The examples include adapting Lil-gp to BOINC (\textbf{Method 1}) and using the wrapper for ECJ (\textbf{Method 2}).

\subsection{Lil-gp and BOINC}
Lilgp is a well known C GP system (see \cite{lilgp}). As Lil-gp is coded in C, porting the framework to BOINC is not difficult (\textbf{Method 1}). 
The main porting changes were done
in all the Input/Output routines that access files. Under BOINC the I/O routines are treated with specific I/O functions due to
the parallel nature of the model. Once the changes were done Lil-gp was ready to be compiled and linked with the BOINC libraries.

In summary, a Lil-gp-BOINC  project is composed by the following items:
\begin{itemize}
    \item \emph{Binary}. The Lil-GP compiled problem (symbolic linear regression, even parity 5, etc.) using the adapted Lil-gp-BOINC
        framework.
    \item \emph{Input files}. Lil-GP uses as input file the GP parameter file (probability of crossover, mutation, etc.).
    \item \emph{WU}. The WU for this project specifies which are the input files needed by the Binary, the output files which are going
        to be generated by the Binary and finally if it is necessary the command line arguments that can be passed to the Binary. In
        this project, it's necessary to specify the input file using a command line argument.
\end{itemize}

The results that we described below show the effectiveness of the approach. Nevertheless, researchers frequently don't have the time to manage
a porting or the tool is written in a language different to C or Fortran. In this cases the goal is to use the tool as it is. Next
sub-section deals with this case.

\subsection{ECJ and BOINC}

ECJ is a modern JAVA framework for Evolutionary Computation (EC) \cite{ecj}. This framework can run different kinds of EC techniques
like: genetic algorithms, evolutionary strategies, genetic programming, etc.

As ECJ is not coded in C++ or Fortran there are two ways of supporting BOINC:
\begin{enumerate}
    \item \emph{Port the framework}. This solution is quite difficult due to the framework is written in JAVA. As is written in a
        different programming language, the port step implies a complete rewriting of all the framework using C++ or Fortran.
    \item \emph{Use the Wrapper}. This is a simple solution consisting of to not modify any source code line to run the framework inside BOINC. 
\end{enumerate}

Therefore we employ the wrapper solution. However this solution implies that all the clients should have installed a JAVA virtual machine,
because without JAVA it is impossible to run any ECJ experiment. So, the adopted solution to support JAVA inside all the clients, was to 
pack also the JAVA virtual machine with ECJ. In summary, the ECJ-BOINC project is composed by the following items:
\begin{itemize}
    \item \emph{Binary}. The binary is the wrapper. The Wrappers uses a file called \emph{job.xml} which specifies the \emph{unmodified
        binary} that must be launched by BOINC, the \emph{unmodified binary} input/output files and the command line arguments.
    \item \emph{Input files}
        \begin{itemize}
            \item \emph{ECJ and JAVA}. ECJ and JAVA are packaged in different compressed files. These files will be unpacked when the clients have downloaded them.
            \item \emph{Unmodified Binary}. The \emph{unmodified binary} is a script file which is in charge of unpacking all the input files (ECJ and JAVA virtual machine)
        and start the execution of ECJ. Additionally it's also in charge of handling the internal ECJ checkpointing in order to
        restart, when necessary, from the last saved and stable stage.
        \end{itemize}
    \item \emph{WU}. The WU for this project specifies which are the input files needed by the Binary and the output files which are going
        to be generated by the Binary (in this case by the \emph{unmodified binary}). 
\end{itemize}

In summary the workflow of ECJ-BOINC, once the clients have downloaded all the necessary files, will be the next: 
\begin{enumerate}
    \item The Wrapper launches the starter script.
    \item The script:
        \begin{enumerate}
            \item Unpacks ECJ and JAVA compressed files.
            \item Checks if an ECJ checkpoint file had been created and in that case it will launch ECJ with the checkpoint file, else
                it will launch ECJ in the normal way.
            \item Copies the ECJ output file to the solution file and exits.
        \end{enumerate}
    \item The wrapper waits until the solution file is created and then notifies to the BOINC core client that an execution has ended and that the
        result files can be uploaded to the BOINC server.
\end{enumerate}

Any other statically linked tool could also follow this scheme to be run on BOINC clients. The next section explains how Lil-gp-BOINC and ECJ-BOINC were 
employed to run different experiments using a geographically distributed pool of clients.
\section{Experiments \& Results}
\label{experiments}
The goal of all the experiments presented before is to show that VGC is a useful computing platform for running GP problems. We are not
interested in checking the quality of obtained results, and therefore we have employed the standard implementation of benchmark problems
provided by the tools Lil-gp and ECJ. We focus on measure the performance improvement that it is possible to achieve when a BOINC model 
is used compared with the traditional and sequential mode of running only one machine. Usually speed up is
measure by the standard equation \ref{formula:acceleration}:
\begin{equation}
    \label{formula:acceleration}
A = \frac{T_{seq}}{T_{B}}
\end{equation} where:
\begin{itemize}
    \item $A$ is the acceleration.
    \item $T_{seq}$ is the consumed time by the sequential mode.
    \item $T_{B}$ is the consumed time by the BOINC mode. 
\end{itemize}
Nevertheless, given the special features of VGC, we also measure the available computing power (CP) by using the method described by Anderson and Fedack in \cite{boinc-power}:
\begin{equation}
    \label{formula:boinc-power}
    CP = X_{arrival}*X_{life}*X_{ncpus}*X_{flops}*X_{eff}*X_{onfrac}*X_{active}*X_{redundancy}*X_{share}
\end{equation}
In all the experiments, $X_{redundancy}$ is equal to 1 because we didn't use the redundancy facility provided by BOINC.
$X_{share}$ is also equal to 1 because none of the clients shared its resources with other BOINC projects. $X_{arrival}$ and
$X_{life}$ are important variables due to they measure the host churn (the volunteer computing project's pool of hosts is dynamic). The
rest of the variables measure specific hardware features \cite{boinc-power}. 

\subsection{Lil-gp-BOINC}
This first experiment presents the proof of concept and was set up on a controlled environment, a laboratory, using 
Lil-gp, \textbf{Method 1} see Section \ref{howto-use-boinc}. In order to measure the performance improvement we chose 
the Artificial Ant on Santa Fe Trail problem (see \cite{koza:book}).

We run 25 executions of the experiment with different population sizes (1000 and 2000 individuals) and
generations (1000 and 2000). Two pools of clients, one with 5 and another with 10 machines, were used for running the Lil-gp-BOINC model.  

As said above, given the aim of this research we don't present the quality of obtained results (which are the same as the sequential
execution). We focus on the performance, computing power and speed up. Table \ref{tab:lilgp} shows the consumed time by Lil-gp-BOINC, 
standard Lil-gp and the acceleration which was obtained using 5 and 10 client machines with the BOINC model.

\begin{table}
    \centering
    \caption{\label{tab:lilgp} Execution time for Lil-gp and Lil-gp-BOINC}
    \subtable[Using 5 Clients]{\begin{tabular}{c|c|c|c|}
       \cline{2-4}
       \cline{2-4} & $ T_{seq} $ & $ T_{B} $ & Acc.\\
       \hline \multicolumn{1}{|c|}{1000 Gen, 1000 Ind.} & 4250s & 1548s & 2.7455  \\
       \hline \multicolumn{1}{|c|}{1000 Gen, 2000 Ind.} & 650s & 395s & 1.6456  \\
       \hline \multicolumn{1}{|c|}{2000 Gen, 1000 Ind} & 9200s & 2356s & 3.9049  \\
       \hline
   \end{tabular}}
   \subtable[Using 10 Clients]{\begin{tabular}{c|c|c|c|}
       \cline{2-4}
       \cline{2-4} & $ T_{seq} $ & $ T_{B} $ & Acc.\\
       \hline \multicolumn{1}{|c|}{1000 Gen, 1000 Ind.} & 4250s & 1033s & 4.1142 \\
       \hline \multicolumn{1}{|c|}{2000 Gen, 1000 Ind} & 9200s & 1623s & 5.6685 \\
       \hline
   \end{tabular}}
\end{table}

From the above results we can conclude that as more clients are used better performance is obtained. For instance, when we are using 10
clients we have achieved an acceleration of 5 while with 5 clients we only get an acceleration of 3 using the same
number of generations and individuals. It is important to point out that this performance grows as more clients collaborate with the project. 
As this experiment is a proof-of-concept the measure variables, speed up and computing power, have not taken into account real
volunteer users. Therefore, we do not show the available computing power obtained in a volunteer computing scenario (see equation
\ref{formula:boinc-power}).

In order to continue the evaluation of the model we decided to face a more complex GP tool besides a more computing demanding GP
problem.

\subsection{ECJ-BOINC}
This second experiment employs a modern and complex JAVA GP framework (ECJ). For this tool we decided to use the wrapper (Method 2) in order to support BOINC.
We chose the GP benchmark Boolean Multiplexer function (see \cite{koza:book}).
In general, the input to the Boolean Multiplexer function consists of $k$ ``address'' bits $a_{i}$ and $2^{k}$ ``data'' bits $d_{i}$
which has the form $a_{k-1}\cdots a_{1}a_{0}d_{2^{k-1}}\cdots d_{1}d_{0}$ with a length equal to $k + 2^k$. The search space for this
function is equal to $2^{k+2^{k}}$. 

\begin{figure}
    \caption{\label{fig:clients}Distributed Infrastructure}
    \begin{center}
        \mbox{
        \subfigure[Interconnected Cities]{\includegraphics[width=7cm]{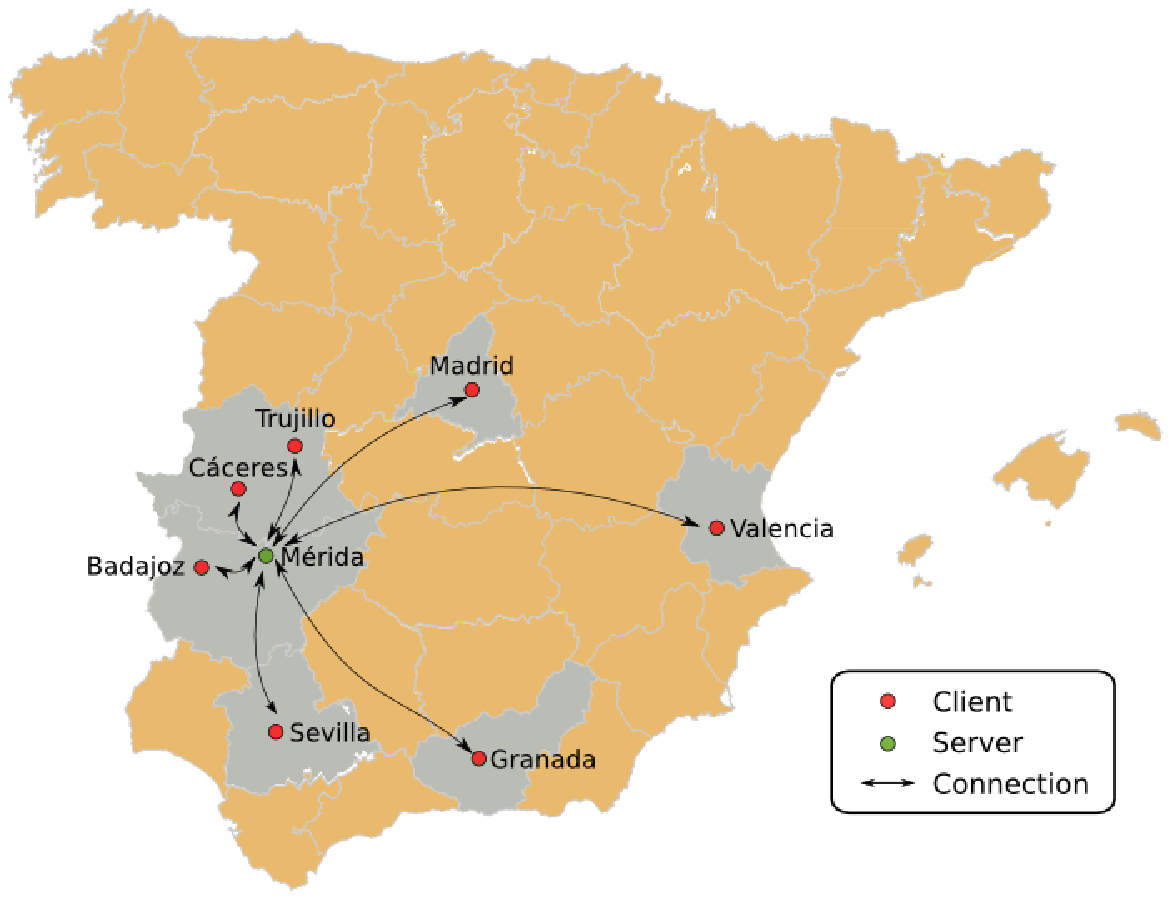}}
        \subfigure[Clients per City]{\includegraphics[width=7cm]{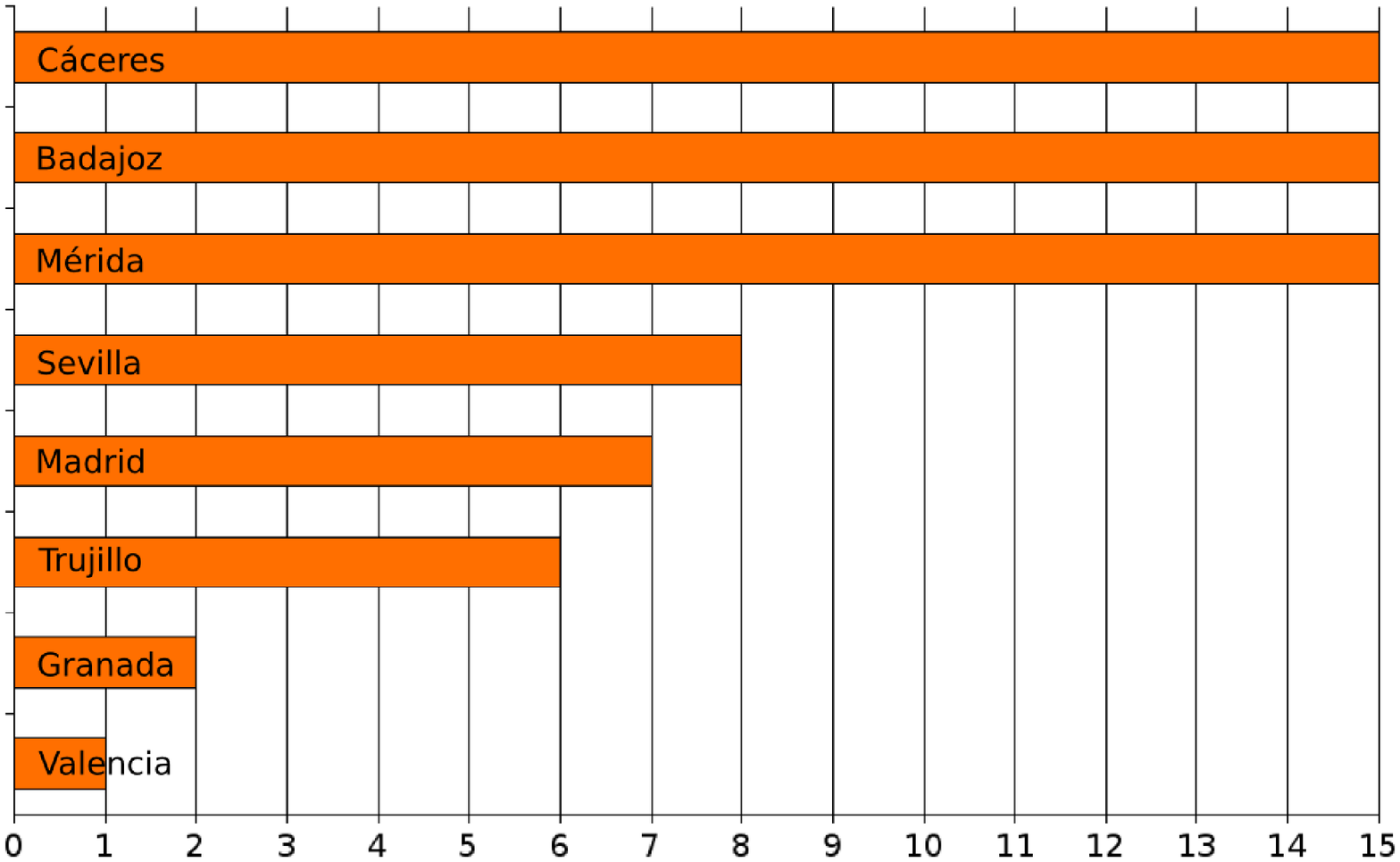}}
        }
    \end{center}
\end{figure}

This problem has been run in several geographically distributed
laboratory clients belonging to the University of Extremadura (Cáceres, Badajoz and Mérida), see Fig. \ref{fig:clients}(a). Fig. \ref{fig:clients}(b) shows the number 
of clients per
city which take part in the experiment. It's important to point out that in following tables, $T_{B}$ measures all the employed
time (client connection, WU download, CPU time, results upload, etc.):
since the first client registers and collaborate with the project until the last connection to the server from any client.

\begin{figure}
    \caption{\label{fig:host-churn}Host churn during September of 2007}
    \begin{center}
        \includegraphics[width=7cm]{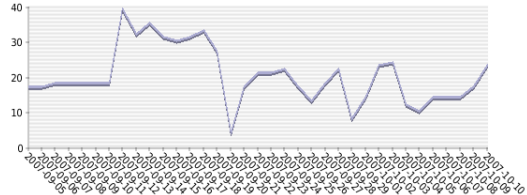}
    \end{center}
\end{figure}

828 iterations of the 11 multiplexer function ($k=3$) were initially performed using 45 computers. 
The experiment used the same Koza parameters (4000 individuals
and 50 generations) for more details see \cite{multiplexer-11}. From the 828 iterations 449 iterations found the perfect solution
(although this is not the goal of our research) to
the 11 multiplexer problem. Some iterations gave an error due to the initial and default restriction of the JAVA heap
size\footnote{The heap size was later modified to avoid this problem.}. 119.18 seconds in average were needed in order to
find the perfect solution while 134.75 seconds in average is the needed time to run one execution. From the 45 available computers,
only 27 produced 828 results. The achieved speed up was 0.29 which means a deterioration in the performance. The reason is the easiness
of the problem, only 134.75 seconds in average, and we have to take into account that $T_{B}$
measures also the host churn (see \ref{fig:host-churn}). Taking into account that the project was running for 5.35 days, $X_{life}$ is measure only from the first
connection to the last communication of hosts that had no communicated in at least one day. Thus, the achieved CP is
equal to  80GigaFLOPS. This CP was obtained because the project was running only a few days and all the hosts were active 
during all the computation time. Tab. \ref{tab:multiplexer}(a) shows a summary.

We increased the complexity of the problem with $k=4$ (compare the new search space $2^{1048576}$ with the previous one $2^{2048}$). Our interest is not in solving the
problem, but in setting up a time consuming experiment for testing the VGC model. We tried first to deploy 42 runs of 
an experiment using 50 generations with a population of 1000 individuals. The rest of the GP parameters are the same as Koza parameters 
for the 11 multiplexer see \cite{multiplexer-11}.

Volunteer computers from other Universities or institutions such as: CICA in Sevilla, University of Extremadura (Cáceres, Badajoz,
Mérida), Granada, Valencia, UNED in Madrid, and Ceta-Ciemat in Trujillo collaborated with the project. Thus, the computing resources 
are more heterogeneous and reallistic now.  Using this infrastructure we performed 42 runs of the experiment.  
A performance improvement was achieved as this problem needs in average 31079.28 seconds to run one
execution. 41 machines were used to solve this problem. From 41 computers, 7 produced the 42 runs due to some machines
were turned off for hours, others still computing, etc. (typical VGC behavior). Thus, the obtained acceleration was 1.95 (see Tab. \ref{tab:multiplexer}). The
obtained speed up was nice although not impressive but it was obtained for free with a quite small number of volunteers involved. 
In average one iteration employs 30944.53 seconds more
than the 11 multiplexer. $X_{life}$ was also measure as in the 11 multiplexer problem due to the project was running only few days. So,
the achieved computing power is equal to 23 GigaFLOPS and is smaller because we are employing only 41 hosts and
because the project was running 7.75 days. However, bear in mind that BOINC has a pool of 2,364,170 available computers which could
collaborate with a project in the future instead of only 42 which we have employed here, not all of them simultaneously available.

Tab. \ref{tab:multiplexer}(b) shows a summary of the relevant data. The best found Fitness was
$Raw=180224.0$ $Adjusted=5.54862e-06$  $Hits=868352.0$.
\begin{table}
    \centering
    \caption{\label{tab:multiplexer} Execution time for ECJ and ECJ-BOINC}
    \begin{tabular}{c|c|c|c|c|}
       \cline{2-5}
       \cline{2-5} & $ T_{seq} $ & $ T_{B} $ & Acc.  & CP\\
       \hline \multicolumn{1}{|c|}{11 bits, 828 runs, 50 Gen, 4000 Ind.} & 134078s & 462259s & 0.29 & 80 GFLOPS \\
       \hline
       \hline \multicolumn{1}{|c|}{20 bits, 42 runs, 50 Gen, 1000 Ind.} & 1305330s & 669759s & 1.95 & 23 GFLOPS \\
       \hline
   \end{tabular}
\end{table}

Finally, we performed another experiment with a complex and not statically linked GP tool which makes impossible to employ Method 1 or
2. Additionally, this experiment faces a real life and time-consuming Computer Vision problem (instead of a benchmark problem) that has already been solved in a sequential 
fashion (Interest Point detectors,see \cite{ipgp}). This GP framework uses the Matlab environment and several image tool-boxes,
which implies a much more complex system, being therefore more difficult to deploy it over a BOINC infrastructure. 
Hence, our proposal is to use a Virtualization layer (see \cite{vmware2}) inside BOINC by creating an image of a working scientific
system ( hardware, OS and the research tool). Thanks to this new virtualization layer (based on VMware \cite{vmware2}) any GP system or framework 
-independently from its complexity, programming language and operating system- can be run on any BOINC client (Linux, Windows or
MacOSX, for further details see \cite{vmware-boinc-ipgp}).

For this experiment we set up 10 Ms Windows volunteer computers. The virtual image was build using a GNU/Linux x86
operating system. Thus, a GNU/Linux scientific environment runs directly inside Ms Windows thanks to the Virtual-BOINC approach. The 10
Windows PCs produced 12 solutions during 48 hours. The consumed time by each solution was in average of 18 
hours. The total time consumed for producing 12 solutions by a sequential run was 215 hours. Therefore, thanks to this new model the obtained speed up was of 4.48 and a CP of
25.67 GFLOPS, (see Tab. \ref{tab:ipgp}).
\begin{table}
    \centering
    \caption{\label{tab:ipgp} Execution time for IP and IP-Virtual-BOINC}
    \begin{tabular}{c|c|c|c|c|}
       \cline{2-5}
       \cline{2-5} & $ T_{seq} $ & $ T_{B} $ & Acc.  & CP\\
       \hline \multicolumn{1}{|c|}{75 Gen, 75 Ind.} & 215h & 48h & 4.48 & 25.67 GFLOPS \\
       \hline
   \end{tabular}
\end{table}

In summary, the BOINC model improves significantly the performances when CPU-intensive, time-consuming, real-life problems are solved by means
of GP. Moreover, as more computers collaborate with one project more computing power and speed
up is achieved with a free cost.
\section{Conclusions}
\label{conclusions}
This paper has presented a new approach to solve GP problems using Volunteer Grid Computing. Three methods have been described
in order to support BOINC. Firstly, porting one easy GP application to BOINC, secondly using a modern and more complex GP framework which was run inside
BOINC without any modification due to it's statically linked and finally a complex GP environment which faces a real-life problem by
using a virtualization layer which allows running inside BOINC any GP system independently of its complexity, programming language or operating
system. Three experiments were performed, one in a controlled environment, another over a geographically distributed 
infrastructure involving 8 cities and finally one more using the virtualization extension.
Results shows that VGC is a perfect springboard to run complex and time-intensive problems by means of GP using free computing resources.
Moreover, BOINC is a really interesting technology if we take into account the big pool of BOINC enabled computers 2,364,170 which could
collaborate with a new GP BOINC project.
\section{Acknowledgments}
This work was supported by Junta de Extremadura, Cátedra CETA-CIEMAT de la Universidad de Extremadura,  Regional Gridex project
PRI06A223 and National NOHNES project TIN2007-68083-C02-01 Spanish Ministry of Science and Education.
\bibliography{../../../Bibliografia/articulos,../../../Bibliografia/enlaces}
\bibliographystyle{splncs}
\end{document}